\documentclass[12pt]{iopart}
\usepackage{harvard}
\usepackage{astrojournals}
\usepackage{epsfig}
\usepackage{setstack}
\usepackage{amssymb}
\usepackage{multirow}
\usepackage{multicol}
\usepackage{color}
\usepackage[varg]{txfonts}
\usepackage{url}
\usepackage{iopams}  

\def \msun{\rm M_\odot}

\begin{document}

\title[SMBH: removing the symmetries]{SMBH accretion \& mergers: removing the symmetries}

\author{Andrew King$^{1,\star}$ \& Chris Nixon$^{1,2,3}$}

\address{1 Department of Physics \& Astronomy, University of
  Leicester, Leicester, LE1 7RH, UK} \address{2 JILA, University of
  Colorado \& NIST, Boulder, CO 80309-0440, USA} \address{3 Einstein
  Fellow} \address{$\star$ ark@leicester.ac.uk}

\begin{abstract}
We review recent progress in studying accretion flows on to supermassive black holes (SMBH). Much of this removes earlier assumptions of  symmetry and regularity, such as aligned and prograde disc rotation. This allows a much richer variety of effects, often because cancellation of angular momentum allows rapid infall. Potential applications include lower SMBH spins allowing faster mass growth and suppressing gravitational--wave reaction recoil in mergers, gas--assisted SMBH mergers, and near--dynamical accretion in galaxy centres.
\end{abstract}

\maketitle

\section{Introduction}
It is now widely accepted that almost all reasonably large galaxies have supermassive black holes (SMBH) in their centres, and further that these holes grow predominantly through luminous accretion of gas \cite{Soltan1982,YT2002}. Any realistic model of gas flows around black holes must take into account the angular momentum of the gas. Radiation processes can relieve the gas of orbital energy, but there is no equivalent process which can intrinsically remove angular momentum. As a result, gas flows tend to form rotationally supported discs with characteristic radii given by the angular momentum of the flow \cite{Pringle1981}. Inward gas flow -- accretion -- would be impossible without some process (usually called `viscosity') transporting angular momentum outwards in the disc, allowing mass to flow inwards \cite{LP1974} and release its gravitational binding energy. Near the black hole this becomes very large, approaching a significant fraction of the rest--mass energy. For this reason, accretion discs probably power the most luminous objects in the universe, and understanding the nature of their angular momentum transport has been a major goal of modern astrophysics. The most promising candidate is the magnetorotational instability (MRI) \cite{BH1991} which injects turbulence into the gas by stretching magnetic field lines.

The viscosity coefficient in accretion discs is often assumed isotropic and parameterised as \cite{SS1973}
\begin{equation}
\label{nu}
\nu = \alpha c_{\rm s} H,
\end{equation}
where $c_{\rm s}$ is the sound speed, $H$ is the disc angular semi-thickness and $\alpha<1$ is a dimensionless parameter. In principle $\alpha$ is a function of position within the disc gas. This formalism characterises the maximum viscosity feasible in an accretion disc through an efficiency parameter: any turbulent gas velocities above the sound speed would quickly shock and dissipate, and any turbulent length scales longer than $H$ would lead to a disc thicker than $H$. The maximum viscosity is then $\approx c_{\rm s}H$ \cite{Pringle1981}. In practice much of the physics of accretion discs depends only on low powers of $\alpha$, allowing simple insights by taking $\alpha$ as a global constant.

From (\ref{nu}) we can write down the accretion timescale $t_{\rm visc} = R^2/\nu$ for a disc as
\begin{equation}
t_{\rm visc} \approx 10^{10}~{\rm
  yrs}~\left(\frac{\alpha}{0.1}\right)^{-1}\left(\frac{H/R}{10^{-3}}\right)^{-2}
\left(\frac{M}{M_8}\right)^{-1/2}\left(\frac{R}{1~{\rm
    pc}}\right)^{3/2}.
\end{equation}
This is evaluated here for typical active galactic nuclei (AGN) disc parameters: $\alpha \sim 0.1$ from \citeasnoun{Kingetal2007}, and $H/R \sim 10^{-3}$ from e.g. \citeasnoun{CD1990}. This form already allows an important conclusion about AGN accretion. From our remarks above, this is how SMBH grow, and so must occur on timescales significantly shorter than the age of the universe. Requiring $t_{\rm visc}$ to be shorter than a Hubble time, we see that the disc scale radius must be $R\ll 1$~pc.

We reach a similar conclusion by considering the effects of self-gravity on an AGN disc. At a radius of only $\sim 0.01$~pc \citeasnoun{Kingetal2008} \citeaffixed{Levin2007,Goodman2003}{see also} find that discs become gravitationally unstable. For SMBH this instability is catastrophic and results in a complete fragmentation. Most of the disc gas forms stars, starving the inner disc \cite{Goodman2003,Levin2007,Kingetal2008}. This suggests that discs feeding SMBH must either be fed at mass rates {\it many} orders of magnitude below Eddington, or form as small scale `shots'. Observations of Eddington accretion suggest the latter.

These simple arguments already point to a new picture for SMBH accretion which we discuss here.

\section{SMBH accretion}
\label{bhdiscs}
Astrophysical black holes have two important parameters: mass and angular momentum (spin). The spin controls the specific binding energy of matter accreting on to the hole, and so the accretion luminosity for a given black hole mass. Accretion grows the SMBH mass, but also affects its spin. 

The huge disparity in scales between the black hole ($R_{\rm g} = 5\times 10^{-6} M_8$~pc) and the galaxy ($\sim 10-100$~kpc) strongly suggests that the angular momenta of the black hole spin and of the gas trying to accrete on to it cannot be correlated in direction, at least initially. In particular the orbital plane of most of the disc mass is unlikely to be aligned with the SMBH spin plane. (This is the first example of several we shall encounter where strong symmetry assumptions [here alignment], originally made for simplicity, turn out to have a distorting effect in suppressing various important effects.)

The first work on the behaviour of misaligned discs was by \citeasnoun{BP1975}. This was actually in the context of stellar--mass black hole binary discs where for example an asymmetric supernova kick may have significantly misaligned the black hole spin and the binary orbit \citeaffixed{Roberts1974}{e.g.}. The physics here is set by the Lense--Thirring effect. \citeasnoun{LT1918} showed that the dragging of inertial frames makes misaligned test particle orbits precess around the angular momentum of a gravitating body at a rate
\begin{equation}
{\bi \Omega}_{\rm p} = \frac{2G{\bi J}_{\rm h}}{c^2R^3},
\end{equation}
where ${\bi J}_{\rm h}$ is the angular momentum (here of the black hole). This precession is strongly differential -- much faster for gas close to the black hole. The precession time is
\begin{equation}
\label{tlt}
t_{\rm LT} = \frac{1}{\left|{\bi \Omega}_{\rm p} \right|} =
\frac{c^2R^3}{2GJ_{\rm h}} = \frac{1}{2a}\left(\frac{R}{R_{\rm
    g}}\right)^2 \frac{R}{c}
\end{equation}
where $J_{\rm h} = \left|{\bi J}_{\rm h}\right| = aGM^2/c$ \cite{KP1985}. From (\ref{tlt}) it is easy to see that for gas orbiting close to the black hole horizon ($R\approx R_{\rm g}$) the precession time can be as short as the dynamical time (at which point the orbit is no longer near-circular). However, the precession is strongly dependent on radius, so for gas far from the black hole the precession is entirely negligible:
\begin{equation}
t_{\rm LT} = 6.5\times 10^{10} a^{-1} \left(\frac{M}{10^8
  \msun}\right)^{-2}\left(\frac{R}{1{\rm pc}}\right)^3 {\rm ~yrs}.
\end{equation}
Thus on scales $\gtrsim 1$~pc, precession induced by the SMBH can be ignored.

\citeasnoun{BP1975} considered a disc of gas subject to a strongly differential precession of this type (modelled in Newtonian gravity, as everything we shall consider). These authors suggested that dissipation in the disc between the differentially precessing rings causes the disc to align to the black hole spin, more quickly in the centre than in the outer parts. So after some time the inner disc is aligned, the middle disc is warped and the outer disc is still misaligned. For an isolated disc-hole system, the warp propagates outwards until the entire disc is aligned. Later, \citeasnoun{PP1983} showed that these early investigations into warped discs \citeaffixed{Petterson1977,Petterson1978,Hatchettetal1981}{e.g.} did not properly handle the internal fluid dynamics in a warped disc. In particular their equations did not fully conserve angular momentum, although the conclusions of \citeasnoun{BP1975} still hold qualitatively. \citeasnoun{PP1983} investigated the Navier-Stokes equations for a warped disc in the linear (small warp) regime. They discovered that the warp can propagate in two distinct ways. In the first of these, viscosity dominates ($\alpha > H/R$), and things behave diffusively, as in \citeasnoun{BP1975}. The second mode of propagation is wave-like, and pressure dominates ($H/R > \alpha$). For black hole discs, the diffusive mode is expected to dominate as the discs are generally thin ($H/R \sim 10^{-3}$) and viscous ($\alpha\approx 0.1$). We therefore focus on this case. For wave-like discs see e.g. \citeasnoun{PL1995}; \citeasnoun{LO2000} \& \citeasnoun{Lubowetal2002}.

Global solutions in full 3D hydrodynamics were hard to achieve, so to make progress \citeasnoun{Pringle1992} used conservation equations to derive an equation governing the evolution of a twisted disc composed of circular rings interacting viscously. Using the equations of \citeasnoun{Pringle1992}, \citeasnoun{SF1996} calculated the secular evolution of warped discs. A subtle error in their calculation led them to conclude that discs always coaligned with the black hole spin -- even if they began close to counteralignment. This led to the conclusion that all black hole discs align on timescales $\sim \alpha^2 t_{\rm visc}$. As $\alpha < 1$, with observations suggesting $\alpha \approx 0.1$ \cite{Kingetal2007}, this disc--BH coalignment occurred long before significant accretion could take place. This largely removed the motivation for studying misaligned or warped discs. 

The alignment error had serious consequences: since alignment would always occur rapidly whatever the initial orientation of disc and hole spin, SMBH would always gain almost all their mass from {\it prograde} accretion discs, thus spinning up to near--maximal values after doubling their masses. This in turn made the specific binding energy release maximal (about 40\% of rest--mass energy). Since radiation pressure inhibits accretion (via the Eddington limit), this implied that only rather low mass accretion rates were possible. The discovery of SMBH with masses $\gtrsim 10^9\msun$ at redshifts $z \gtrsim 6$ (allowing accretion only for a timescale of order $10^9$~yr) then appeared to require that these holes must have started from initial `seeds' with significant masses $\gtrsim 10^6\msun$.

Almost a decade passed (with the belief in rapid coalignment well entrenched in SMBH models) before \citeasnoun{Kingetal2005} discovered the error in \citeasnoun{SF1996} - the implicit assumption of infinite disc angular momentum. With this restriction lifted \citeasnoun{Kingetal2005} showed that counteralignment simply requires
\begin{equation}
\label{crit}
\cos\theta < -\frac{J_{\rm d}}{2J_{\rm h}}.
\end{equation}
where $\theta$ is the angle between the disc and hole angular momentum vectors with magnitude $J_{\rm d}$ \& $J_{\rm h}$ respectively. This condition ensures that the the total angular momentum vector (sum of disc and hole) is {\it shorter} than the hole angular momentum. Since the latter is subject only to precessions, its length cannot change during the alignment process, meaning that hole and disc must end up opposed to account for the shorter total angular momentum.

With retrograde discs now plausible -- indeed likely, \citeasnoun{KP2006} \& \citeasnoun{KP2007} began to explore the effect of misaligned discs on the evolution of SMBH. The larger lever--arm of retrograde accretion flows near the black hole horizon offers an obvious way of keeping their spins low, and so making SMBH growth efficient, particularly if accretion disc events have no preferred direction with respect to the host galaxy. This scenario \citeaffixed{KP2006,KP2007}{often called `chaotic accretion';} is consistent with many observational effects, such as the observed lack of correlation of the directions of AGN jets (thought to be orthogonal to the plane of the accretion disc close to the hole) with large--scale properties of the host \cite{Kingetal2008}. There was renewed interest in warp propagation through accretion discs. \citeasnoun{LP2006} used the numerical method of \citeasnoun{Pringle1992} to explore the evolution predicted by \citeasnoun{Kingetal2005}, confirming that the criterion (\ref{crit}) correctly determines the alignment history (co, counter, or more complex) of accreting black holes.

In the chaotic accretion picture, a natural question is what happens when a misaligned accretion event occurs on to a pre-existing disc. \citeasnoun{Nixonetal2012a} explored this with numerical simulations. As partially opposed gas flows interact through viscous spreading, significant amounts of angular momentum can be cancelled, leading to gas infall and strong accretion.

Perhaps surprisingly, this cancellation can occur even in a single disc event, as sufficiently inclined discs may change their planes almost discontinuously, rather than in the smooth warps envisaged by BP \cite{NK2012}. There had been sporadic evidence that discs could `break' in this way \cite{Larwoodetal1996,FN2010,LP2010}. The theoretical possibility of this idea was further studied by \citeasnoun{Ogilvie1999}: for a locally isotropic viscosity coefficient, he showed that the stresses in a strongly warped disc would evolve in such a way that the forces trying to bring the disc back into a single plane would actually weaken as the warp grew. 

The first systematic investigation of this possibility was by \citeasnoun{NK2012}, who used the \citeasnoun{Pringle1992} method to explore warp propagation with the full effective viscosity coefficients derived by \citeasnoun{Ogilvie1999}. This revealed modified Bardeen--Petterson (BP) behaviour where a sharp break in the disc occurs between the aligned inner disc and the misaligned outer disc. However, the method of \citeasnoun{Pringle1992} forces the disc to respond viscously (following a diffusion equation) to the Lense-Thirring precession and excludes other possible hydrodynamical effects. The sharp disc break found in \citeasnoun{NK2012} suggested that this assumption was too restrictive, and that a full 3D hydro numerical approach was needed.

Accordingly \citeasnoun{Nixonetal2012b} made SPH (smoothed particle hydrodynamics) simulations of misaligned discs around a spinning black hole. SPH was already known to reproduce the behaviour derived by Ogilvie in modelling the communication of a warp in a fluid disc \cite{LP2010}, and therefore suitable for modelling such discs. \citeasnoun{Nixonetal2012b} confirmed the expected BP behaviour of low--inclination discs, but showed that misaligned discs can break. Further, a break can promote cancellation of disc angular momentum. Separated disc annuli precess independently of each other, and so inevitably become partially opposed. In effect each annulus borrows angular momentum from the central black hole to achieve the cancellation, so that in some way the black hole is complicit in feeding itself more rapidly than simple viscous evolution would allow.

\begin{figure}
  \begin{center}
    \includegraphics[angle=0,width=\columnwidth]{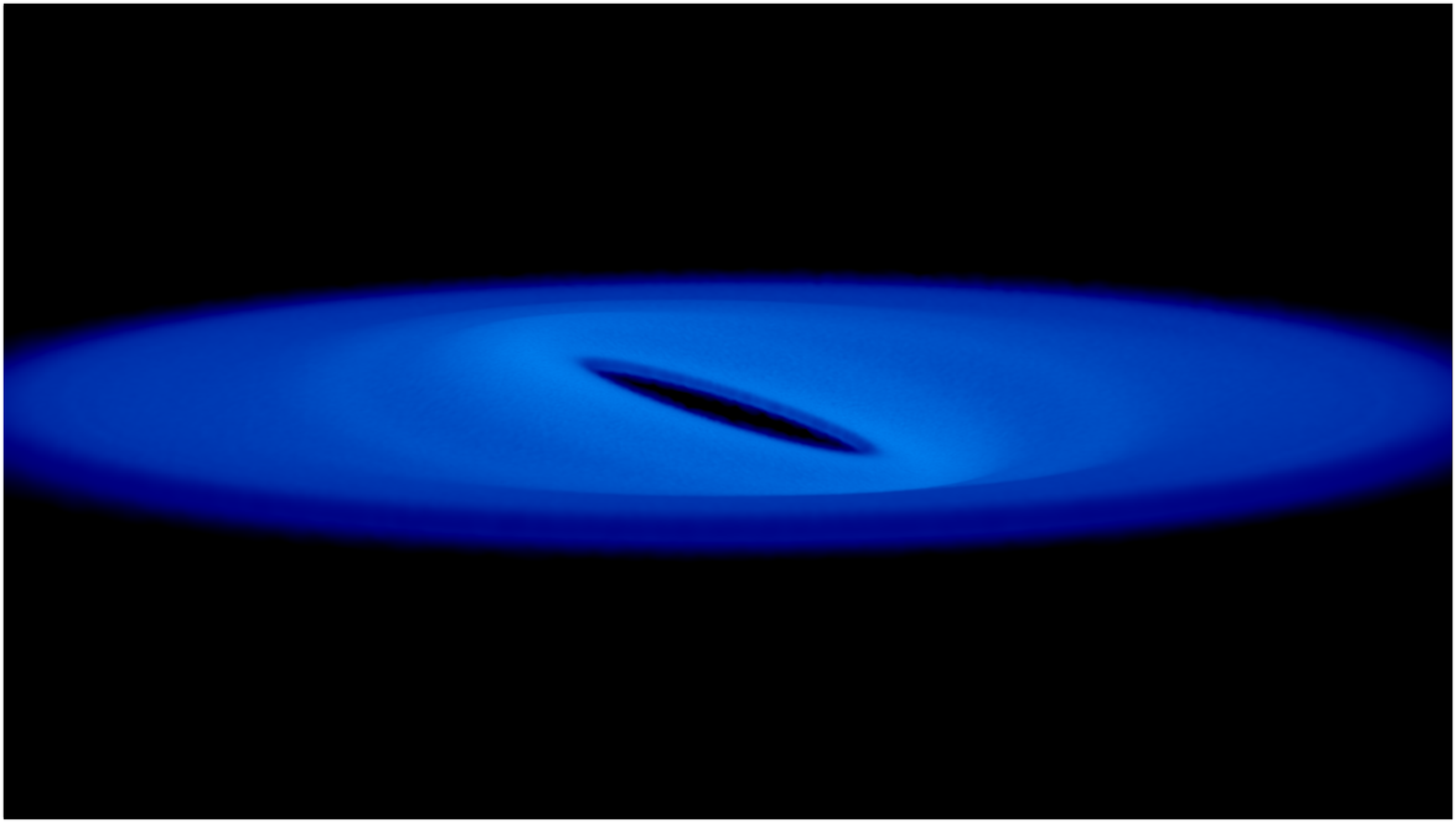}
    \includegraphics[angle=0,width=\columnwidth]{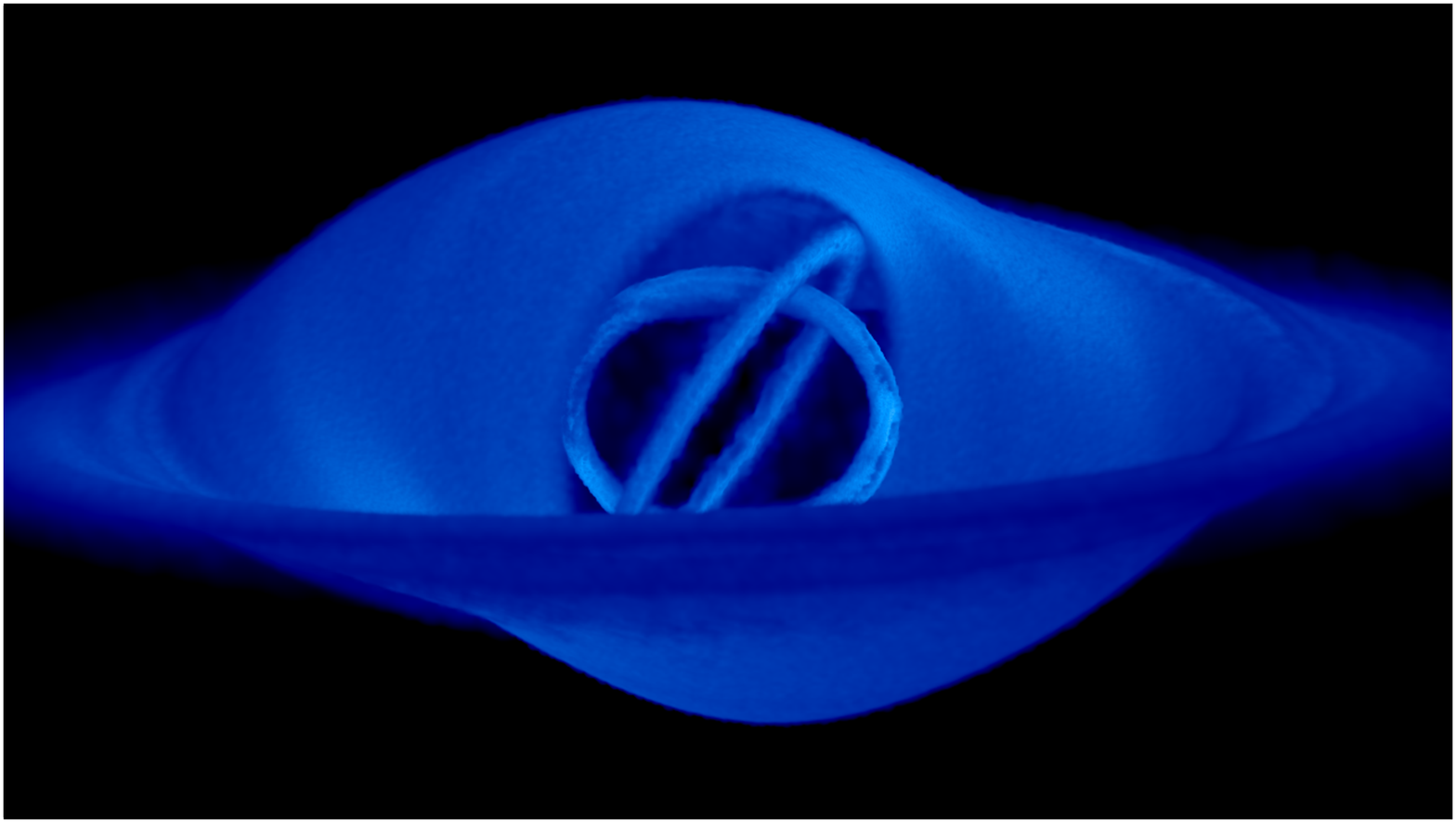}
    \caption{3D disc structures of the small (top) and large (bottom) inclination simulations from \citeasnoun{Nixonetal2012b}.}
    \label{bh}
  \end{center}
\end{figure}

This tearing behaviour differs radically from smooth BP evolution, and we are only beginning to understand its implications. Fig.~\ref{bh} shows the 3D disc structure for a small and large inclination disc around a spinning black hole \cite{Nixonetal2012b}.

Importantly, it is clear that although tearing was first noted for accretion on to spinning black holes, the crucial element is that disc orbits precess differentially. This always happens if the effective potential for the disc flow has a quadrupole component. We can therefore expect tearing to occur quite generically for disclike flows on all scales, not merely around spinning black holes, but also around SMBH binaries (see below) and on the scale of an entire galaxy.

The potential objection to the ideas of breaking and tearing is that current treatments rely on a scalar viscosity (stress proportional to strain), whereas a more general tensor relation might conspire to hold the vertical disc structure together. Intuitively this seems unlikely for the currently--favoured MRI picture: where the `horizontal' viscosity driving angular momentum transport and accretion is constantly pumped by azimuthal winding up of magnetic fields, the vertical relative motion of  the two sides of a warp is periodic and bounded, suggesting that if anything an MRI viscosity might be still weaker in the vertical direction. Current MRI simulations are very far from being able to answer such questions, not least because of the unphysical diffusive effects of having a disc plane inclined with respect to the grid symmetries.

\section{SMBH binaries}
\label{binaries}
Most galaxies have supermassive black holes, and galaxy mergers are common. Dynamical friction drives the SMBHs of a merging galaxy pair close together in the centre of the merged galaxy: binary angular momentum is lost to surrounding stars. This forms an SMBH binary with a typical separation of order 1 pc, but cannot coalesce the holes, since all the suitable stars have been driven away by the dynamical friction process itself. Since very few SMBH binaries are actually observed, some as yet unknown process must drive the binary to coalescence. This is the last parsec problem \cite{Begelmanetal1980,MM2001}.

There have been various suggested solutions, generally of two types. One type invokes collisionless matter to arrange that stellar orbits passing close to the SMBH binary are constantly refilled, and so available to remove it angular momentum. Recent attempts along these lines have met with some success, e.g. with triaxial dark matter (DM) haloes refilling the stellar orbits \cite{Bercziketal2006}, or using the Kozai mechanism in triple SMBH systems \cite{Blaesetal2002}, or non-axisymmetric potentials \cite{Iwasawaetal2011}. However it seems clear that at some level gas must play a role, and attempts to invoke it constitute the second type of proposed solution of the last parsec problem. Shocks in the galaxy gas flows can rob the gas of the angular momentum supporting its orbit so that it falls towards the binary with a random inclination. Quite separately, to observe SMBH binaries in action we need to understand how they interact with gas.

All early studies of gas interacting with an SMBH binary were, as in Section~\ref{bhdiscs}, limited to discs which were both coplanar and prograde wrt the binary. But then two physical effects severely limit the effect on binary evolution. First, the disc mass is limited by self-gravity to 
\begin{equation}
\label{mdisc}
M_{\rm d} \lesssim \frac{H}{R}M_{\rm b},
\end{equation}
where $H/R \ll 1$. So any disc with enough mass to affect the binary quickly fragments into stars. Second, an infinite family of prograde disc orbits are resonant with the binary rotation, holding the gas far out. In principle this shrinks the binary by transferring angular momentum to the gas, but much too slowly to be helpful. Rather as in the case of dynamical friction, the effect which removes angular momentum from the binary tends to weaken itself by gradually pushing away the agency (here the prograde disc gas) which effects this removal in the first place.

But there is no compelling reason to suppose that the circumbinary gas disc is prograde. And the reasoning of the last paragraph strongly suggests that things would work much better with retrograde discs, as happened with SMBH growth in Section~\ref{bhdiscs}. Retrograde discs do not suffer resonances \cite{PP1977}, so disc gas can instead accrete freely on to the binary with {\it negative} angular momentum. \citeasnoun{Nixonetal2011a} show that once a retrograde moving mass  $\sim M_2$ has interacted with the binary, its eccentricity approaches unity (here `interacted' means gravitationally - the gas need not accrete for example). Even before this happens, gravitational wave losses complete the SMBH coalescence. So the timescale for this phase of the merger is given by $\sim M_2/{\dot M}$, where ${\dot M}$ is the accretion rate through the retrograde disc.

However, (\ref{mdisc}) tells us that any individual accretion event cannot have a mass $\gtrsim M_2$ unless the mass ratio is very small ($\lesssim H/R$). Therefore we must consider multiple, randomly--oriented events. As prefigured at the end of the last Section, \citeasnoun{Nixonetal2011b} showed that much of the disc alignment and breaking phenomena derived for a disc around a single spinning black hole hold for a circumbinary disc. Here the quadrupole part of the binary potential induces a similar (but stronger) precession to the Lense-Thirring effect. In particular \citeasnoun{Nixonetal2011b} conclude that counter--alignment of the circumbinary disc occurs if and only if
\begin{equation}
\cos\theta < -\frac{J_{\rm d}}{2J_{\rm b}}
\end{equation}
exactly as (\ref{crit}) for the black hole case. \citeasnoun{Nixon2012} showed that counteralignment of a circumbinary disc is stable (contrary to previous reports), and that for any reasonable set of parameters the binary dominates the angular momentum of the system. So approximately half of all randomly oriented accretion events are retrograde.

All this means that a sequence of $\sim 2q R/H$ randomly oriented accretion events, each limited by self--gravity, can drive the binary eccentricity close to unity and allow gravitational wave losses to complete the black hole merger. But there is a subtlety here. If we assume that collisionless processes can drive the binary in to 0.1 pc, the viscous time (appropriate for retrograde circumbinary discs) is
\begin{equation}
\label{tnu}
t_{\rm visc} \approx 4.5\times 10^8 {\rm ~yr~} \left(\frac{\alpha}{0.1}\right)^{-1}\left(\frac{H/R}{10^{-3}}\right)^{-2}\left(\frac{M}{10^8\msun}\right)^{-1/2}\left(\frac{R}{0.1{\rm pc}}\right)^{3/2}.
\end{equation}
This time is so long that at first glance it appears unlikely that gas accretion can help on these scales. The merger time is given by
\begin{equation}
\label{tm}
t_{\rm merge} \sim 2q\frac{R}{H} t_{\rm visc}
\end{equation}
where taking $t_{\rm visc}$ from (\ref{tnu}) gives an upper limit. For typical numbers, $q R/H \approx 10-100$, and so (\ref{tm}) would be longer than a Hubble time.

\begin{figure*}
  \begin{center}
    \includegraphics[angle=0,width=\textwidth]{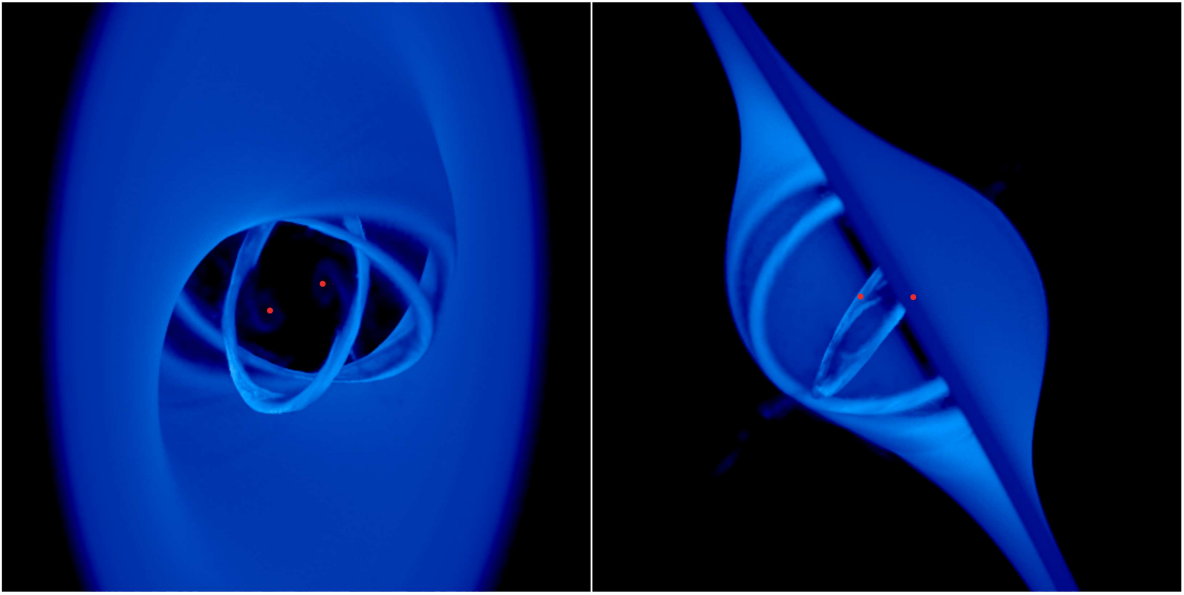}
    \caption{The 3D structure of the $\theta=60^\circ$ circumbinary disc from \citeasnoun{Nixonetal2013}.}
    \label{bintear}
  \end{center}
\end{figure*}

\begin{figure}
  \begin{center}
    \includegraphics[angle=0,width=\columnwidth]{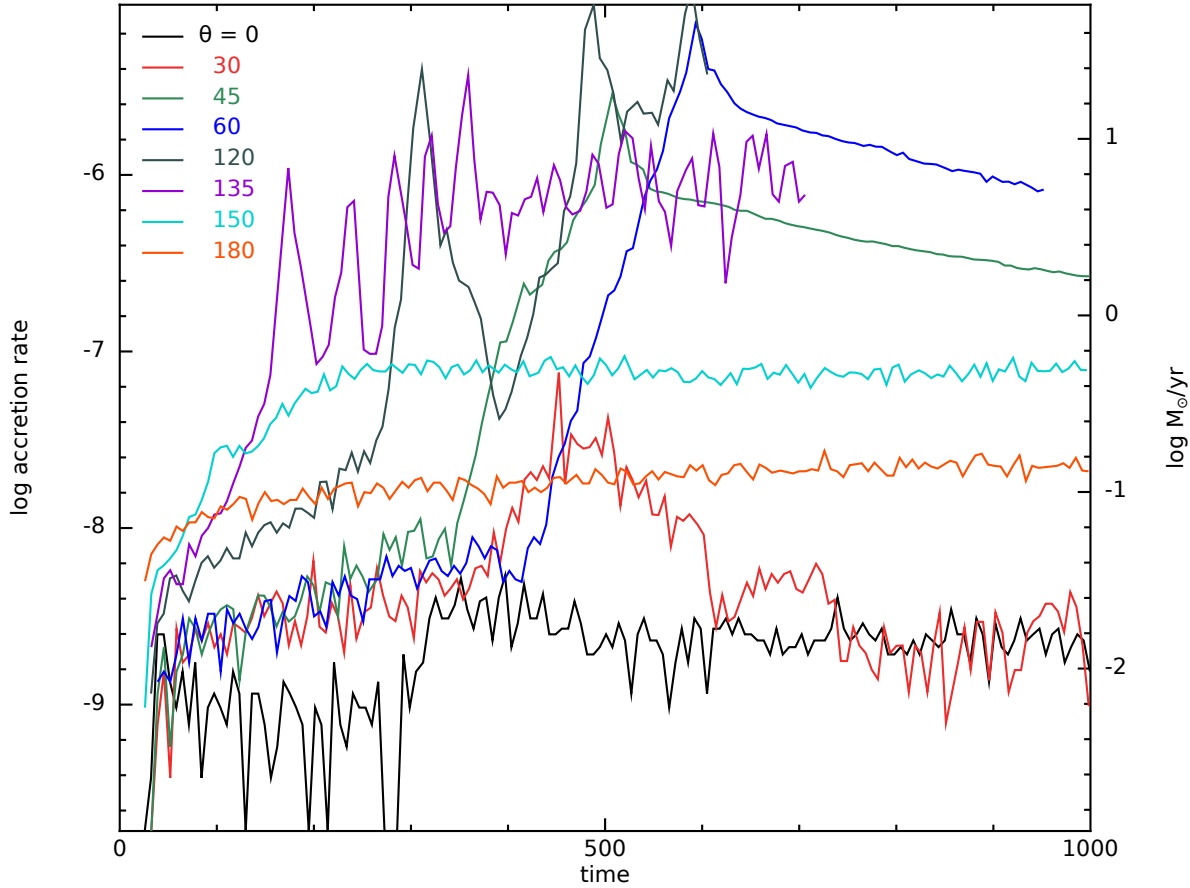}
    \caption{Accretion rates for the circumbinary disc simulations in \citeasnoun{Nixonetal2013}. For tearing discs the accretion rate can be boosted by factors up to $10^4$ compared to the equivalent prograde, aligned disc.}
    \label{binacc}
  \end{center}
\end{figure}

However \cite{Nixonetal2013} recently showed that misaligned circumbinary discs can tear and cancel angular momentum (see Fig.~\ref{bintear}), and therefore the accretion rates may be increased by factors up to $10^4$ for sustained periods (see Fig.~\ref{binacc}). So depending on the mass supply from the galaxy, it is possible that the gas--driven merger timescale can be as short as $\sim 10^6-10^7$~yr from 0.1 pc, and considerably shorter still if collisionless dynamics puts the starting point even further in.


It appears then that collisionless \citeaffixed{Bercziketal2006,Khanetal2011}{e.g.} and collisional processes (e.g. those considered here) conspire together to solve the last parsec problem with collisionless processes driving the binary to the scales ($\lesssim 0.1$~pc) where gas can have a strong effect. Then, if collisionless processes can go no further, gas completes the merger provided the galaxy can supply enough mass ($\sim M_2$) on the right (chaotic) orbits.

\section{Conclusion}
\label{con}

The common thread in most of the work reported here is the gradual removal of assumptions of regularity and symmetry of gas flows near supermassive black holes. These assumptions included those of disc flows aligned with the spin plane of a single black hole, or the orbital plane of an SMBH binary, as well as prograde rotation in most cases. Although probably initially made on grounds of simplicity, these assumptions have little basis in reality. We have seen that they arbitrarily rule out a rich variety of phenomena and actually create artificial difficulties in many cases. In particular by keeping the sense of rotation of everything parallel, they make angular momentum barriers to infall and mergers formidable. The central regions of galaxies are not in general strongly coherently rotating, so cancellations between opposed gas flows must be a common phenomenon. So we have seen that in general SMBH do not inevitably have high spins -- indeed if accretion events have no preferred direction there is a slow but persistent statistical trend towards lower spins \cite{Kingetal2008}, something that was already known to be the outcome of repeated SMBH coalescences \cite{HB2003}. Several problems are greatly eased by this -- we have already noted that the Eddington limit of a slowly spinning hole is much less of a barrier to black hole mass growth, and in addition note that low spins make strongly anisotropic gravitational wave reaction unlikely, allowing galaxies to retain the merged black holes. For completeness we should add that the same physics makes the often--invoked idea of jet precessions seem unlikely \cite{NK2013}. This paper also points out that SMBH spins barely move under the effect of individual accretion events, and instead perform very slow random walks in direction. 

Perhaps the most spectacular consequence of removing symmetry assumptions is that disc flows may generically be subject to breaking and tearing \cite{NK2012,Nixonetal2012b,Nixonetal2013}. By temporarily borrowing angular momentum from the source of the potential, gas rings arrange to cancel it with neighbours to allow accretion to occur on near--dynamical timescales. This may potentially add considerable new insight to our current picture of many gas--dynamical effects in astrophysics.

\ack
Research in theoretical astrophysics at Leicester is supported by an STFC Consolidated Grant. CN acknowledges support for this work, provided by NASA through the Einstein Fellowship Program, grant PF2-130098. 

\section*{References}
\bibliographystyle{jphysicsB} 
\bibliography{nixon}

\end{document}